\documentclass{article}

\usepackage{graphicx}

\begin{document}

\title{\bf Food-chain competition influences gene's size.} 
\date{}
\maketitle

\author{Marta Dembska$^1$, Miros{\l}aw R. Dudek$^1$\footnotetext[1]{mdudek@proton.if.uz.zgora.pl} and Dietrich Stauffer$^2$\footnotetext[2]{stauffer@thp.Uni-Koeln.DE}}

\vspace*{0.2cm}
$^1$ { \it Institute of Physics, Zielona G{\'o}ra University,
 65-069 Zielona G{\'o}ra, Poland}\\
$^2$ {\it Institute of Theoretical Physics, Cologne University, D-50923 K{\"o}ln, Euroland}
~\\
~\\
~\\
~\\
\begin{abstract}
We have analysed an effect of the Bak-Sneppen predator-prey food-chain self-organization on nucleotide content of evolving species. In our model, genomes of  the species under consideration have been represented by their nucleotide genomic fraction and we have applied two-parameter Kimura model of substitutions to include the changes of the fraction in time. The initial nucleotide fraction and substitution rates were decided with the help of random number generator. Deviation of the  genomic nucleotide fraction from its equilibrium value was playing  
the role of the fitness parameter, $B$, in Bak-Sneppen model. 
Our finding is, that the higher is the value of the threshold fitness, during the evolution course, the more frequent are large fluctuations in  number of species with strongly differentiated nucleotide content; and 
it is more often the case that the oldest species, which survive the  food-chain competition,
might have specific nucleotide fraction making possible  generating long genes. 
 
\end{abstract}



\section{Model introduction}

To understand the way the higher organized species emerge 
during evolution we consider very simple model of evolving food 
chain consisting of $N$ species. In the model, each species is represented by
nucleotide composition of their DNA sequence and the substitution rates between the nucleotides. There are four possible nucleotides, A, T, G, C, 
in a DNA sequence. In our model, the DNA sequence is represented, simply, by  four reals, $F_A, F_T, F_G, F_C$, being the nucleotide fractions and

\begin{equation}
F_A+F_T+F_G+F_C=1.
\label{fractions}
\end{equation}

\noindent
The nucleotide fractions depend on time due to mutations and selection.

Our model is originating from the Bak-Sneppen  model of co-evolution \cite{BS1} and Kimura's neutral mutation hypothesis (\cite{Motoo_Kimura1},\cite{Wen-HsiungLi1}).
According to Kimura's hypothesis, neutral mutations are responsible for  molecular diversity of species.
In 1980, Kimura introduced  two-parameter model \cite{Motoo_Kimura2}, \cite{Wen-HsiungLi2}, where the transitional substitution rate 
(substitutions $A \leftrightarrow G$ and $C \leftrightarrow T$) is different from the transversional rate (substitutions $A \leftrightarrow T$, $G \leftrightarrow T$, $A \leftrightarrow C$, $G \leftrightarrow C$) . If we use  Markov chain notation, with discrete time $t$, then the transition matrix, ${\bf M}_{\rm nucl}$, 

\begin{eqnarray}
  {\bf M}_{\rm nucl} &=& \left(
 \begin{array}{llll}
  1-uW_{A} & u~W_{AT} & u~W_{AG} & u~W_{AC} \\
  u~W_{TA} & 1-uW_{T} & u~W_{TG} & u~W_{TC}\\
  u~W_{GA} & u~W_{GT} & 1-uW_{G} & u~W_{GC}\\
  u~W_{CA} & u~W_{CT} & u~W_{CG} & 1-uW_{C} 
 \end{array}
\right)\\
&=& 
\left(
\begin{array}{llll}
  1-u(2v+s)& uv& us & uv \\
  uv & 1-u(2v+s) & uv & us\\
  us & uv & 1-u(2v+s) & uv\\
  uv & us & uv & 1-u(2v+s) \\
 \end{array}
\right),
\label{macierz1}
\nonumber
\end{eqnarray}

\noindent
representing rates of nucleotide substitutions in the two-parameter Kimura model
fulfills the following equation

\begin{equation}
{\overrightarrow F(t+1)} = {\bf M}_{\rm nucl} {\overrightarrow F(t)}
\label{evolution}
\end{equation}

\noindent
where  {\overrightarrow {F(t)}=\{$F_A(t),F_T(t),F_G(t),F_C(t)\}^T$}  
denotes nucleotide fractions at time $t$, $u$ represents substitution rate and  the
symbols $W_{ij}=s$ for transitions and $W_{ij}=v$ for transversions ($i,j=A,T,G,C$) represent relative
substitution probability of nucleotide $j$ by
nucleotide $i$. $W_{ij}$ satisfy the equation

\begin{equation}
\sum_{i,j=A,T,G,C} W_{ij}=1, 
\end{equation}

\noindent
which in the case of the two-parameter Kimura model is converted into the following

\begin{equation}
4s+8v=1,
\label{suma}
\end{equation}

\noindent
and $W_j=\sum_{i\neq j} W_{ij}$.

\begin{figure}
\begin{center}
\includegraphics[scale=0.4]{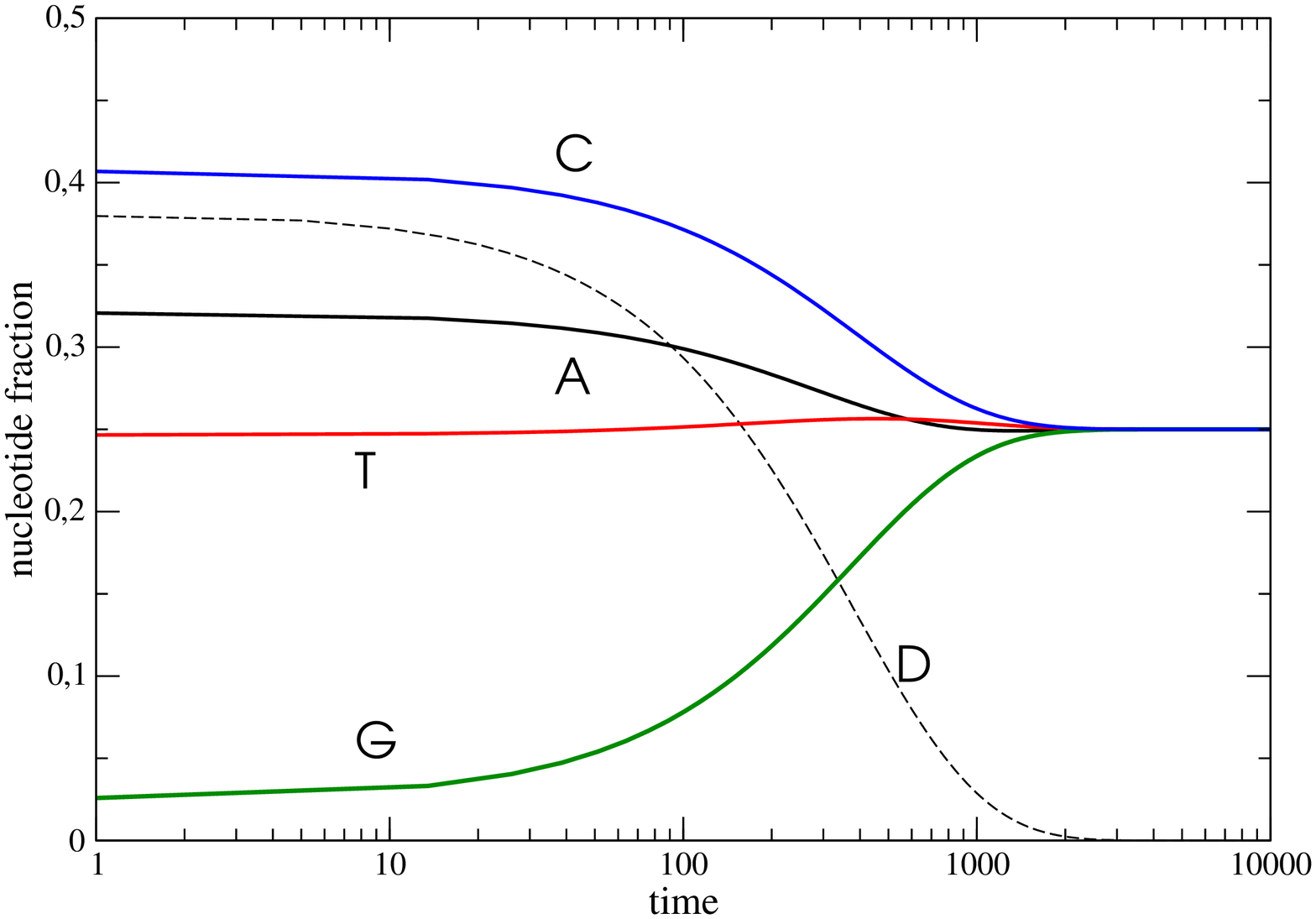}
\end{center}
\caption{Dependence of nucleotide fractions on time in two-parameter Kimura model. Here, the initial fractions take the following values: $F_A=0.320964$, $F_T=0.246541$, $F_G=0.0252434$, $F_C=0.407252$. Besides, there has been plotted the maximum absolute deviation from the difference $\vert F_A-F_T \vert$ and
$\vert F_G-F_C \vert$ (the dashed curve).}
\label{fig1}
\end{figure}

Evolution described by Eq.(\ref{macierz1}) has the property that starting from some initial value of $\overrightarrow F(t_0)$ at $t=t_0$  the solution tends to an equilibrium in which $F_A=F_T=F_G=F_C=0.25$. The example of this type of behavior has been presented in Fig.\ref{fig1}. The two-parameter Kimura approximation is one of the simplest models of nucleotide substitutions. For example, in reconstructing the phylogenetic trees, one should use a more general form of  the transition matrix in Eq.(\ref{macierz1}) (\cite{Wen-HsiungLi2},\cite{Lobry1},\cite{Rzhetsky},\cite{Lobry2}). This  is not necessary in our model, where we need only the property that the nucleotide frequencies are evolving to their equilibrium values.

More complicated prey-predator relations were simulated with a $5 \times 5$ 
Chowdhury lattice \cite{Stauffer2} with a fixed number of six food levels. Each
lower (prey) level contains twice as many possible species as the upper 
(predator) level. Also this model does not contain an explicit bit-string as 
genome. We now 
introduced a composition vector \overrightarrow {F(t)} as above, different for
each different species, and let it evolve according to Eq.(3). Again, after
many iterations all four fractions approached 0.25. This result, as we will show below, is qualitatively different from that  in the model defined below, where we observe fluctuations of nucleotide frequency, instead.

Our model consists of $N$ species and for each species we define  the set of random parameters, {$F_A$, $F_T$, $F_G$, $F_C$, $u$, $s$, $v$}, which satisfy 
only two equations, Eq.(\ref{fractions}) and Eq.(\ref{suma}), and we assume that $4s>8v$ to fulfill the condition that transitions ($s$) dominate transversions ($v$). 
The nucleotide fractions, representing each species, change in time according to Eq.(\ref{evolution}).

The species are related according to food-chain. In the case of the nearest-neighbor relation the species $i+1$ preys on species $i$. The extension to further neighbors follows the same manner.
The food-chain has the same dynamics as in Bak-Sneppen model (BS) \cite{BS1}, i.e.,
 every discrete time step $t$, we choose the species $i$ with minimum fitness $B_i$ and the chosen species  is replaced by a new one together with the species linked to it with respect to food-chain relation. In the original BS model the nearest neighborhood of species $i$ is symmetrical, e.g. $\{B_{i-1},B_i,B_{i+1}\}$. The asymmetrical (directional) neighborhood applied for food-chain modeling has been discussed by Stauffer and Jan \cite{Stauffer1} and their results were qualitatively the same as in the BS model. The generalizations of food-chain onto more complex food-web structures are also available \cite{Ito}, \cite{Stauffer2}.

\begin{figure}
\begin{center}
\includegraphics[scale=0.4]{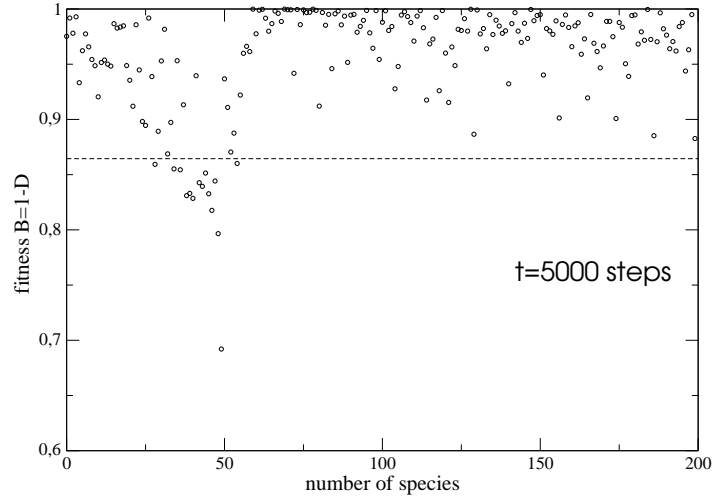}
\end{center}
\caption{Snapshot of the distribution of the species fitness at the transient time $t=5000$. In the example, $N=200$ and the substitution rate is a random real $u=0.01*rnd$, number of the nearest-neighbors $n=1$. The horizontal line  is representing the value of threshold fitness.}
\label{fig2}
\end{figure}

The new species, substituting the old ones, obtain new random values {$F_A$, $F_T$, $F_G$, $F_C$, $u$, $s$, $v$}. In our model  the fitness $B_i$ of the species $i=1, 2, \ldots, N$ is represented by the parameter

\begin{equation}
B_i=1-D, \quad D=\max (\vert {F_A-F_T}\vert, \vert {F_G-F_C}\vert), 
\label{selectionrule}
\end{equation}

\noindent
where $B_i \in [0,1]$ is a measure of the deviation from equilibrium
of the nucleotide numbers $F_A-F_T$ and $F_G-F_C$. Thus, the species with the smallest value of $B_i$ (largest compositional deviation from equilibrium) are eliminated 
together with their right-hand neighbors with respect to food-chain. This elimination mechanism leads to self-organization. Namely, in the case of finite value of $N$
the statistically stationary distribution of the values of $B_i$ ($i=1, 2, \ldots, N$)
is achieved after finite number of time steps with the property that  the selected species with the minimum value $B_{min}$ is always below some threshold value $B_c$ or it is equal to the value. The typical snapshot, at transient time, of the distribution of the values of $B_i$ is presented in Fig.\ref{fig2}. 

 So, if Fig.\ref{fig2} looks much the same as it had been resulted from the simulation of pure BS model, 
then what are the new results in our model?  In the following, we will show that 
the higher value of the threshold fitness, during the evolution course, it is often the case that the winners of the food-chain competition become also species with specific nucleotide composition, which is  generating long genes.

\section{Discussion of results}
We know, from Eq.(\ref{evolution}) (see also Fig.\ref{fig1}), that a single species tends to posses equilibrium nucleotide 
composition, which in this simple two-parameter Kimura model means asymptotically the same nucleotide composition $F_A=F_T=F_G=F_C=0.25$.
 The only distinction, which we could observe, if we had used a more general form of the substitution table, could be the resulting equilibrium nucleotide composition different from the uniform one. This would bring nothing new to the qualitative behavior of our model. 

\begin{figure}
\begin{center}
\includegraphics[scale=0.4]{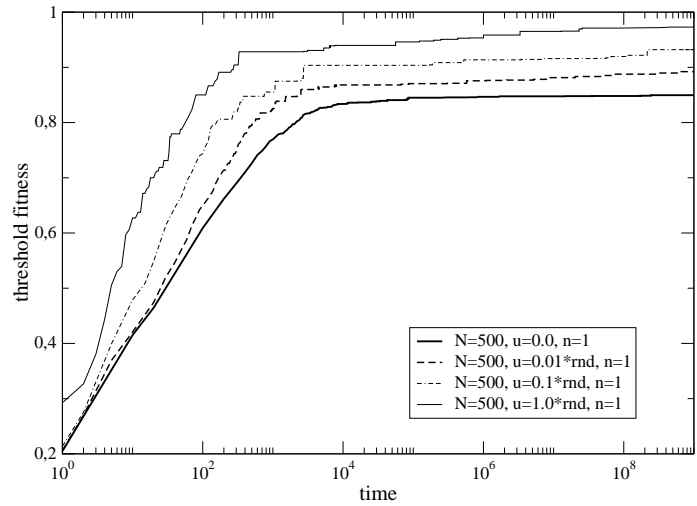}
\end{center}
\caption{Few examples of time dependence of the threshold fitness $B$  for different values of the upper bound of the applied substitution rate $u$.}
\label{fig3}
\end{figure}

\begin{figure}
\begin{center}
\includegraphics[scale=0.4]{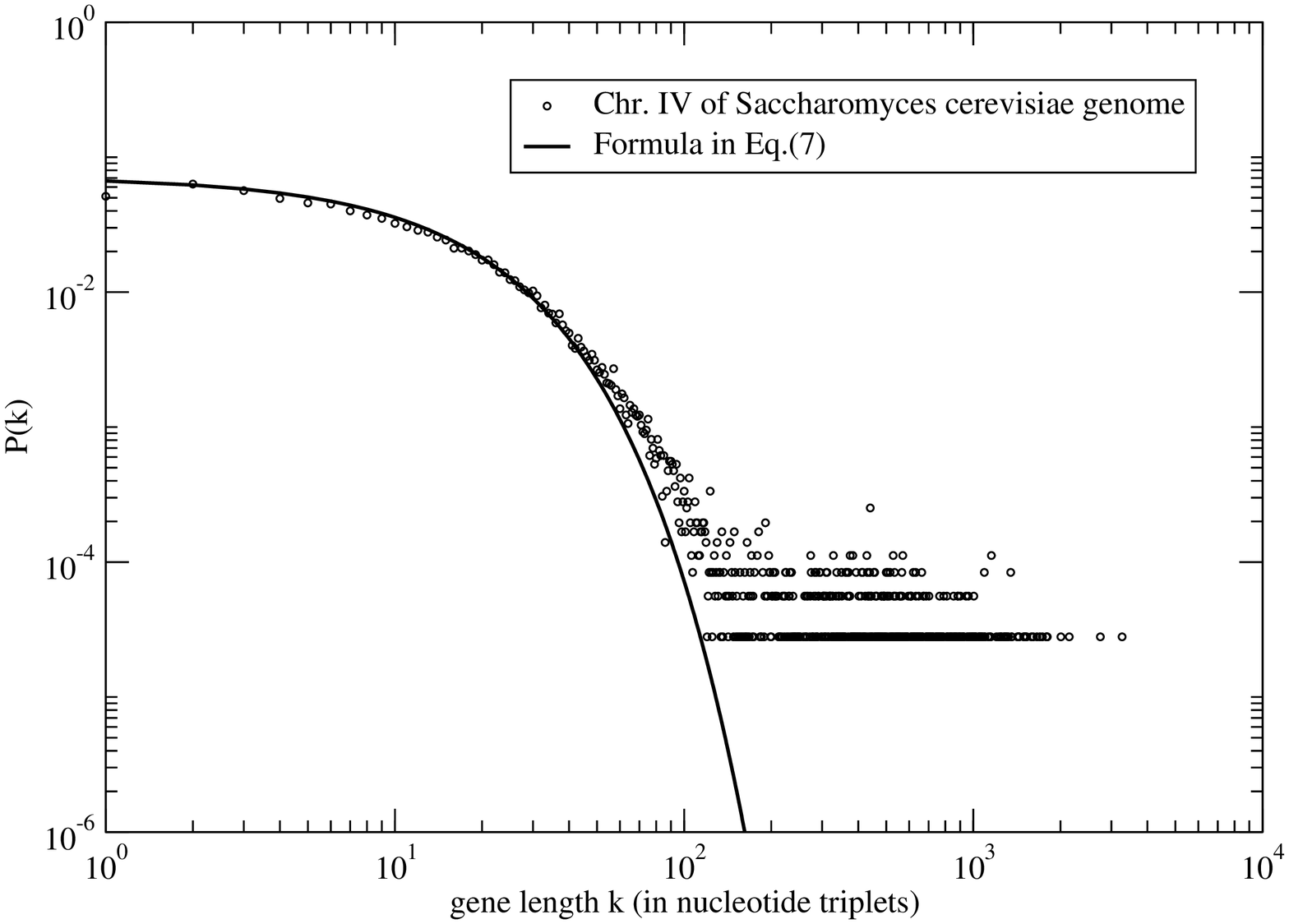}
\end{center}
\caption{
Distribution $P(k)$ of gene length $k$ in Chromosome IV of {\it Saccharomyces cerevisiae} genome and in the case of the approximate analytic  formula (Eq.(\ref{sizeofgene})), where the nucleotide fractions take the values as in Chr. IV, i.e., $F_A=0.31121$, $F_T=0.309727$,  $F_G=0.190188$,  $F_C=0.188875$. Parameter $k$ is representing number of codons (nucleotide triplets).}
\label{fig4}
\end{figure}

Once, in the model under consideration, nucleotide composition of  species is changing  according to Eq.(\ref{macierz1}), the species fitness $B_i$ depends on time. It is not the case in the BS model \cite{BS1}, where the fitness of the evolving species is constant in time unless it is extincted. Although $B_i$ depends on time, the food-chain selection rule introduces mechanism, which  forbids to achieve the equilibrium nucleotide composition ($B_i=1$). Instead, there appears a threshold value of $B_c$, below which the species become extinct. In our model the threshold value  depends on substitution rate $u$. The examples of this dependence for transient time of $10^9$ generations have been plotted in Fig.\ref{fig3}. 

Similarly, as in BS model, the SOC phenomenon disappears if the number of nearest neighbors $n=0$. Then, all species tend to the state with $B=1$.

We will discuss the influence of  threshold fitness optimization on nucleotide composition of species and, in consequence, its influence on the possible maximum length of gene in species genome.
 To this aim, we assume that a gene has continuous structure (no introns) and it always starts from codon START (ATG) and ends with codon STOP (TGA, TAG or TAA).  Then, the probability of generating any gene consisting of $k$ nucleotide triplets  in a random genome with the fractions $F_A$, $F_T$, $F_G$, $F_C$ could be approximated by the following formulae (see also \cite{Cebrat}):

\begin{equation}
P(k)=\alpha F_A F_T F_G (2 F_A F_T F_G+F_A^2F_T ) (1-2 F_A F_T F_G-F_A^2F_T)^{k-1},
\label{sizeofgene}
\end{equation}

\noindent
where $\alpha$ is a normalization constant, which can be derived from the normalization condition

\begin{equation}
\sum_{k=1}^{k_{\rm cutoff}} P(k) = 1.
\label{normalization}
\end{equation}

\noindent
The value of $k_{\rm cutoff}$ in Eq.(\ref{normalization}) could be associated with genome size.

In Fig.\ref{fig4}, there has been shown the relation between the empirical distribution of gene length $k$ in chromosome IV of {\it Saccharomyces cerevisiae} genome and the  distribution $P(k)$ in  Eq.(\ref{sizeofgene}). Similar results we could obtain for other genomes. 
One can observe, that the approximation in Eq.(\ref{sizeofgene}) is acceptable for small gene size, whereas  it becomes wrong for large gene size.  Generally, it is accepted that there is direct selective pressure on gene size for the effect. Examples of papers discussing the problem could be found \cite{WentianLi},\cite{Cebrat},\cite{proteomesize} together with analyses of rich experimental data.

The lowest frequency of gene size, $k$, in  {\it Saccharomyces cerevisiae} genome is equal to $P_0 \approx 2.8 \times 10^{-5}$ (Fig.\ref{fig4}). 
In many natural genomes $P_0$ takes value of the same order of magnitude, e.g., in the {\it B.burgdorferi genome} 
$P_0 \approx 5.7 \times 10^{-5}$. In our model, we have assumed that for all species holds $P_0=1 \times 10^{-6}$.
We have also introduced maximum gene length,  $k_{\max}$, which  is the largest value of $k$ for which $P(k) \ge P_0$.

In the particular case of the same fractions of nucleotides in genome  ($F_A=F_T=F_G=F_C=0.25$) the limiting value 
$k=k_{\max}$ for which $P(k) \ge P_0$ is equal to $k_{\max}=225$ nucleotide triplets ($675$ nucleotides). Thus, in our model, we could expect that for the oldest species the maximum gene length $k_{\max}$ should not exceed the value of $675$ nucleotides. The reason for that is that ageing species  should approach equilibrium composition (Fig.\ref{fig1}). However, surprisingly, we
found that the self-organization phenomenon enforces a  state, in which the oldest species  may have much longer gene sizes than in genome with nucleotide composition corresponding to equilibrium composition. Actually, there start to appear  fluctuations in the number of species with very short genes and very long ones.

\begin{figure}
\begin{center}
\includegraphics[scale=0.4]{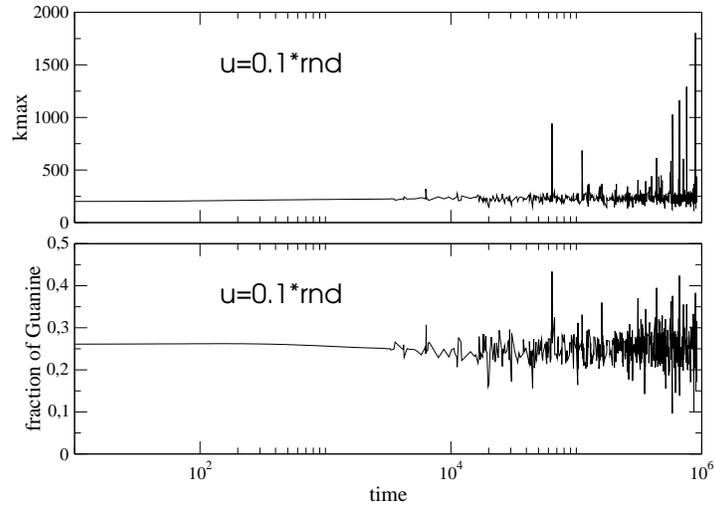}
\end{center}
\caption{Time dependence of  maximum gene size, $k_{\max}$, of the oldest species and the Guanine content in their genome in the evolving ecosystem when $N=500$, $u=0.1*rnd$, $n=1$. The data in the figure have been decimated.}
\label{fig5}
\end{figure}

\begin{figure}
\begin{center}
\includegraphics[scale=0.4]{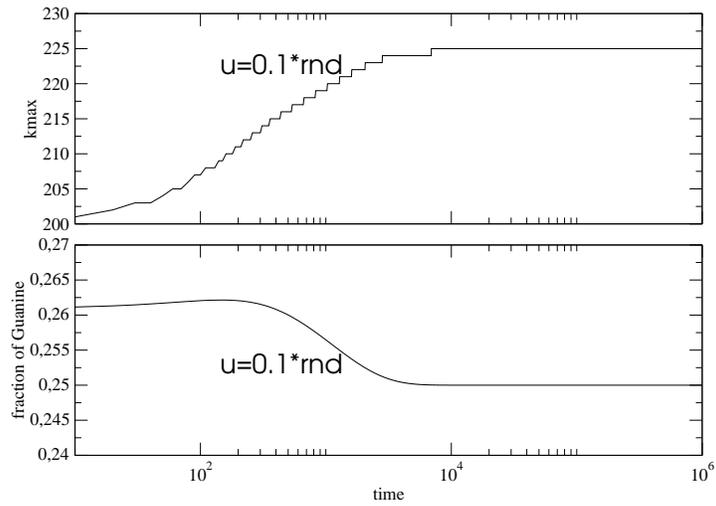}
\end{center}
\caption{The same parameters as in Fig.\ref{fig5} but $n=0$.}
\label{fig6}
\end{figure}

\begin{figure}
\begin{center}
\includegraphics[scale=0.4]{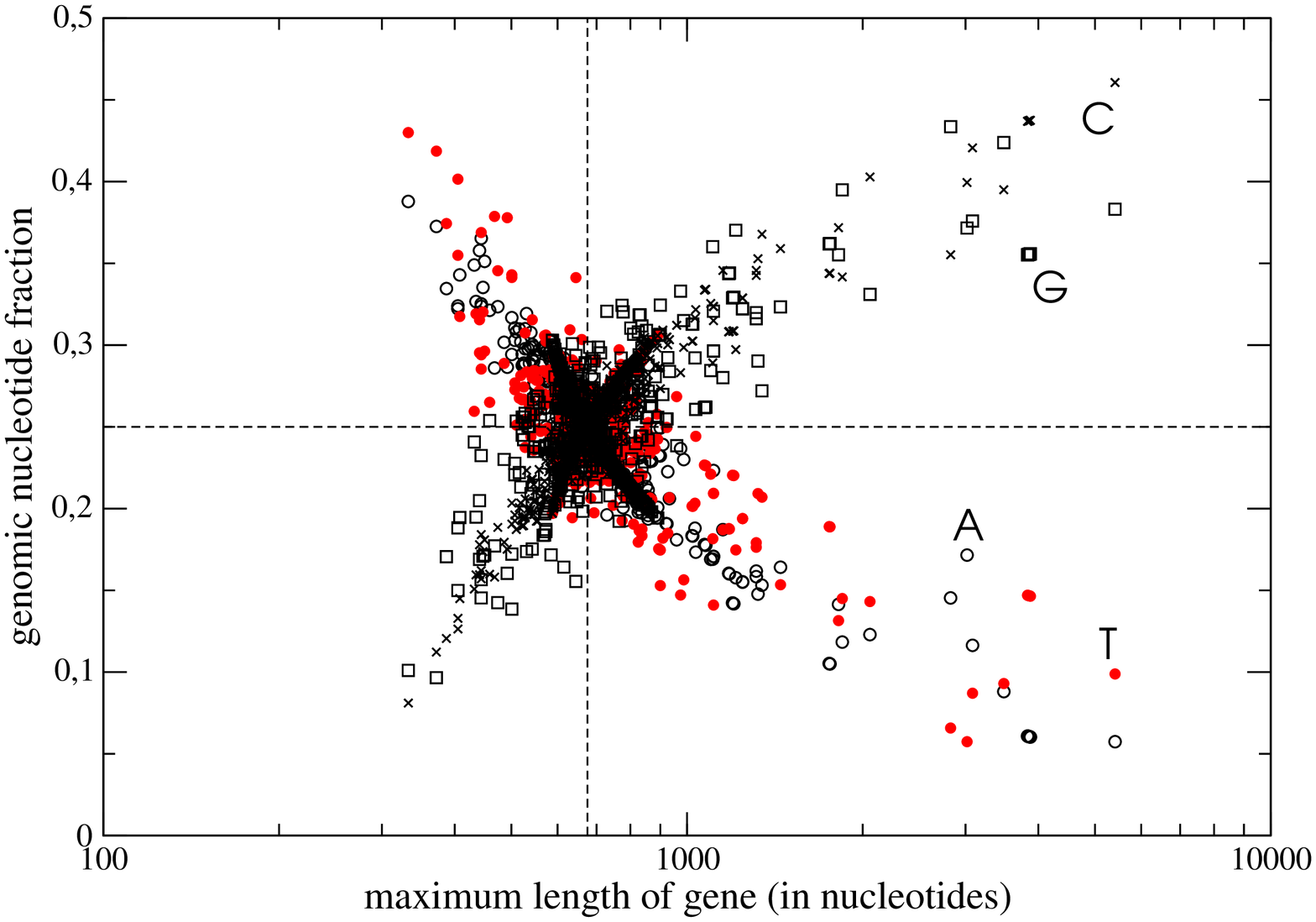}
\end{center}
\caption{Maximum gene length,  $k_{\max}$, versus genomic fraction of nucleotide  in the oldest species. The vertical lines correspond to equilibrium genome. In our model this means  $F_A=F_T=F_G=F_C=0.25$. The same parameters as in Fig.\ref{fig5} have been used.}
\label{fig7}
\end{figure}

The selection towards the species with the smallest deviation from equilibrium nucleotide composition (the largest value of $B$) implicates that the species, which survive the selection, may have specific bias in nucleotide composition, which makes possible  generating long genes. In our model, we have observed abundances of G+C content in the species with long genes. During simulation run, in each time step $t$, we have collected in a file data representing age of the oldest species, the corresponding gene size $k$ and nucleotide frequency. We have observed, that the closer the species fitness $B_i$ is to the threshold fitness the older might be the species and also the species might posses longer genes in its genome. There is no such  effect in the case, when $n=0$. Even if there could appear, at some early time interval, a tendency to generate longer genes, this property would have disappeared after longer evolution time of the system of $N$ species. In Fig.\ref{fig5}, we have plotted time dependence of the recorded maximum length of gene in the oldest species and the corresponding Guanine fraction. One can compare this figure with Fig.\ref{fig6}, where there are no prey-predator relation in the ecosystem ($n=0$). In the latter case, the system is ageing in accordance with the
Eq.(\ref{evolution}) and $B_i \rightarrow 1$ ($i=1, 2, \ldots, N$) and self-organization has not been observed.

The observed property of the competing species has an analogy 
with the behavior of the model of  evolution of evolving legal rules in Anglo-American court, introduced by Yee \cite{KentonKYee} (see Fig. 3 in his paper).

The relation between nucleotide fraction of genome and the possible maximum length of gene in such genome has been shown in the histogram in Fig.\ref{fig7}. The presented data  address, solely, the oldest species. Notice, $A \approx T$ and $G \approx C$ for genomes both with short genes and long ones, whereas $A \approx T \approx G \approx C \approx 0.25$ for genomes with nucleotide composition near equilibrium understood in terms of the  two-parameter Kimura model.
We should remember, that the substitution table for the two-parameter Kimura model (Eq.\ref{macierz1}) is a symmetric matrix and the observed compositional asymmetry results directly from the predator-prey self-organization phenomenon. The right-hand wings, evident in the structure in Fig.\ref{fig7}, do vanish in the case when $n=0$ in spite of the same fitness parameter in Eq.\ref{selectionrule} applied for selection.

 We have not included strand structure in species genome, in our model, since it is represented only with the help of nucleotide fraction. Lobry and Sueoka, in their paper \cite{LobrySueoka}, concluded that if there are no bias in mutation and selection between two DNA strands, then it is expected $A \approx T$ and $G \approx C$ within each strand, and that the observed variation in G+C content between species is an effect of another phenomenon than, simply, asymmetric mutation pressure. Here, we have shown, that such compositional fluctuations of genome could result from ecosystem optimization - no direct selection on genes length  is present in our model. 

The predator-prey rule, in the model under consideration, introduces large fluctuations in nucleotide frequency in the ageing ecosystem, if it is sufficiently old. However, we have not observed this frequency phenomenon in modeling speciation on the Chowdhury lattice \cite{Stauffer2}, as we stated in the beginning. After we have introduced a small change of our model, in such a way, that new species arising in the speciation process were taken always from among the survived species, and  we only slightly were modifying their nucleotide frequency by introducing $d\%$ of changes in their values, then the observed by us fluctuations ceased to exist in the limit $d \rightarrow 0$, as found in the Chowdhury model.

\section{Conclusions}
The specific result of the food-chain self-organization of the competing 
species is that the oldest survivors of the competition might posses strong compositional bias in nucleotides, the abundance of G+C content. In our model, this resulting asymmetry makes possible generating long genes. There was no direct selection applied on the gene length, in the  model. The fluctuation in number of species with long genes and short genes represents rather undirectional noise, 
the amplitude of which is increasing while the ecosystem is ageing. 
The effect ceases to exist if there is no species competition. The same is if we allow only $d\%$ changes of nucleotide frequency in the new formed species, in the limit $d \rightarrow 0$.
 
It could be, that the observed self-organization is an attribute of genes in genome evolution.  Typically, many genes are coupled together in genome in a hierarchical dynamical structure, which resembles complex food-web structure. Some genes may be duplicated but also you can observe fusion of genes or even genomes.

\vspace*{0.2cm}
{\bf Acknowledgments}\\
We thank geneticist S. Cebrat for bringing us physicists together at a 
GIACS/COST meeting, September 2005.
 One of us, M.D., thanks A. Nowicka for useful discussion.

\end{document}